\documentclass[runningheads]{llncs}
\usepackage{graphicx}
\usepackage{physics}
\usepackage{amsmath}
\usepackage{amssymb}
\usepackage{stmaryrd}
\usepackage{xcolor}
\usepackage{hyperref}
\usepackage{algpseudocode}
\usepackage{algorithm}
\usepackage{booktabs}
\usepackage{multirow}
\usepackage{comment}
\usepackage{bm}

\newcommand{\Per}[1]{\text{Per} \; #1}
\newcommand{\supp}[1]{\text{supp}(#1)}

\title{Fast and memory efficient strong simulation of noisy adaptive linear optical circuits}

\author{Timothée Goubault de Brugière \inst{1} \and Nicolas Heurtel \inst{1,2}}
\institute{Quandela, 7 Rue Léonard de Vinci, 91300 Massy, France \and Université Paris-Saclay, CNRS, ENS Paris-Saclay, Inria, Laboratoire Méthodes Formelles, 91190,
Gif-sur-Yvette, France}

\titlerunning{Fast and memory efficient strong simulation of linear optical circuits}
\begin{document}
\let\oldaddcontentsline\addcontentsline
\def\addcontentsline#1#2#3{}
\maketitle
\def\addcontentsline#1#2#3{\oldaddcontentsline{#1}{#2}{#3}}        
\begin{abstract}
Exactly computing the full output distribution of linear optical circuits remains a challenge, as existing methods are either time-efficient but memory-intensive or memory-efficient but slow. Moreover, any realistic simulation must account for noise, and any viable quantum computing scheme based on linear optics requires feedforward. In this paper, we propose an algorithm that models the output amplitudes as partial derivatives of a multivariate polynomial. The algorithm explores the lattice of all intermediate partial derivatives, where each derivative is used to compute more efficiently ones with higher degree. In terms of memory, storing one path from the root to the leaves is sufficient to iterate over all amplitudes and requires only $2^n$ elements, as opposed to $\binom{n+m-1}{n}$ for the fastest state of the art method. 
This approach effectively balances the time-memory trade-off while extending to both noisy and feedforward scenarios with negligible cost. To the best of our knowledge, this is the first approach in the literature to meet all these requirements. We demonstrate how this method enables the simulation of systems that were previously out of reach, while providing a concrete implementation and complexity analysis.

\end{abstract}

\setcounter{tocdepth}{2} 
\tableofcontents

\clearpage

\section{Introduction}

Linear optics stands out as a natural and viable platform for carrying quantum information, enabling applications in secure communication~\cite{bennett1984cryptography,bennett2014cryptography} and quantum networks~\cite{kimble2008quantuminternet,wehner2018quantuminternet}. More importantly, linear optics with feedforward --- the adaptive process of using outcomes from one stage to dynamically adjust the operations performed at a later stage in the circuit --- has been shown to be sufficient for implementing any quantum algorithm~\cite{knill2001klm}. The versatility and computational power of linear optics have led to extensive studies both in the context of passive implementations~\cite{aaronson2011bosonsampling,peruzzo2014vqe,mezher2023solving} and adaptive approaches for quantum applications~\cite{chabaud2021qmladaptive,hoch2025qmladapativeBS} and scalable quantum computing schemes~\cite{yoran2003deterministic,nielsen2004optical,browne2005resource,bartolucci2023fbqc,degliniasty2024spoqc,xanadu2025scalingphotonic}.

To effectively study and characterize optical processes, it is crucial to have efficient algorithms for computing output probabilities. A well-studied case is the computation of a single probability~\cite{scheel2004permanents,aaronson2011bosonsampling}, which can be computed with the permanent of a matrix with repeated rows~\cite{shchesnovich2013asymptotic,aaronson2014generalizing,shchesnovich2020classical}, where efficient and optimized implementations exist~\cite{gupt2019walrus,heurtel2023perceval,seron2024juliapackage}. Though in many cases, the focus isn't limited to a single probability but extends to multiple probabilities or even the entire output distribution. For an input of $n$ photons over $m$ modes, this distribution consists of $\binom{n+m-1}{n}$ possible output states. Computing the full distribution by calculating the permanent independently for each output would be too time-consuming, making it essential to optimize computation time.

Several methods have been proposed to efficiently and exactly compute multiple probabilities~\cite{heurtel2023slos} by storing intermediary results, therefore gaining time by using an intense use of memory. While the proposed methods offer a significant time advantage, the bottleneck now becomes the memory, making the computation quickly intractable on any device with reasonable memory requirements. A more memory-efficient approach would be to iterate through all output states without storing large intermediate results. Moreover, linear optics proves especially useful in scenarios involving feedforward. However, to the best of the authors' knowledge, no efficient algorithm has yet been proposed for strongly simulating linear optical circuits with feedforward. Coupled with the realistic requirement of handling noise~\cite{garciapatron2019simulatingboson,oh2023classicalsimulation,shchesnovich2019noiseboson} in physical experiments, we therefore seek an approach that optimally balances the time-space trade-off and is particularly suited for both feedforward and noisy simulations.

In this paper, we introduce LO-SLAP (Linear Optical Simulation through LAttice of Polynomials), a memory-efficient algorithm that can iterate over the output amplitudes and be extended to feedfoward and noisy simulations with negligible cost. The iteration is achieved by exploiting the fact that amplitudes can also be viewed as nth-order partial derivatives of a multivariate polynomial. By storing information about a lower-order partial derivative, it is possible to calculate all child derivatives more efficiently. The algorithm uses $2^n$ in memory, compared with $\binom{n+m-1}{n}$ for SLOS method of~\cite{heurtel2023slos}, such that we removed the dependency in the number of modes $m$. Even though the memory requirement is still exponential in the number of photons, we can theoretically go as high as $30$ photons on a laptop. We provide a theoretical analysis of the complexity of LO-SLAP and show that we actually increase significantly the problem sizes one can strongly simulate with reasonable computation power.

We explicit how LO-SLAP extends to the noisy and feedforward case:
\begin{enumerate}
\item All intermediate coefficients that are calculated during the algorithm are actually output amplitudes of the same experiment but with input states with a lower photon count. We will show that LO-SLAP naturally computes all possible outputs of any input state with $n$ photons or less. Therefore, \textit{at no extra cost}, LO-SLAP can be used for strong simulation dealing with loss as well. We also show how to incorporate other sources of noise such as distinguishability.
\item Our method also handles feedforward with the only extra computation of computing the updated matrix of the experiment after a measurement --- cost that any algorithm simulating feedforward has to compute. To the best of our knowledge, this is the first time that a strong simulation method handling feedforward has been proposed for linear optics. We compare it against an extended version of SLOS~\cite{heurtel2023slos} and demonstrate a theoretical as well as a practical advantage.
\end{enumerate}

The plan of the article is the following. First in Section~\ref{sec::background} we present the polynomial formalism of linear optical experiments, with a review of the existing methods to perform classical simulations. Then we present LO-SLAP in Section~\ref{sec::slos_tree}, we explicit the data structure, how we update it and how we iterate over the output amplitudes. In Section~\ref{sec::extensions} we present our two extensions: feedforward and noisy simulation. We conclude in Section~\ref{sec::conclu}.

\section{Background and state of the art} \label{sec::background}

\subsection{Formalism of linear optics}
\subsubsection{Fock states.}
We consider a system of $n$ photons and $m$ modes. The canonical states of the system are the so-called Fock states 

\[ \ket{s} = \ket{s_1, s_2, \hdots, s_m}, \; \; \; \; \; |s| = \sum_{i=1}^m s_i = n \]

which give the possible distributions of the photons into the different modes (we read $s_i$ photons in the i-th mode). There are $\binom{n+m-1}{n}$ ways to divide $n$ photons into $m$ modes (see the ``stars and bars'' theorem, for instance in \cite{feller1968introduction}). We note $\Phi_{m,n}$ the set of all $\binom{n+m-1}{n}$ Fock states. 

We also naturally associate a Fock state $\ket{s}$ with an array $r_s \in [m]^n$ that gives the output mode of each photon. For instance, with $\ket{s} = \ket{0,2,1,0,1}$, we can have $r_s = [2,2,3,5]$ or any other permutation of $r_s$.

\

The state $\ket{\psi}$ of a linear optical system is a complex normalized linear combination of all possible Fock states of $\Phi_{m,n}$: 

\[ \ket{\psi} = \sum_{s \in \Phi_{m,n}} \alpha_s \ket{s}, \; \; \; \sum_{s} |\alpha_s|^2 = 1 \]

where each $|\alpha_s|^2$ gives the probability of measuring $s$ if we measure the state $\ket{\psi}$. The coefficients $\alpha_s$ are called the amplitudes of the state.

\subsubsection{Creation and annihilation operators.}

We use the formalism of creation and annihilation operators to represent and manipulate Fock states. We note $a_k^{\dag}$, resp. $a_k$, the creation operator, resp. annihilation operator, on mode $k$. These operators act on Fock states as follows: 
\begin{itemize} 
\item[$\bullet$] for creation operators:
\[ a_k^{\dag} \ket{s_1, s_2, \hdots, s_k, \hdots, s_m} = \sqrt{s_k+1} \ket{s_1, s_2, \hdots, s_k+1, \hdots, s_m}, \]

\item[$\bullet$] for annihilation operators:
\[ a_k \ket{s_1, s_2, \hdots, s_k+1, \hdots, s_m} = \sqrt{s_k+1} \ket{s_1, s_2, \hdots, s_k, \hdots, s_m}. \]

\end{itemize}

We can always rewrite the state of our system as the action of a linear combination of products of creation operators on the vacuum state: 
\[ \ket{\psi} = \sum_{s \in \Phi_{m,n}} \alpha_s \ket{s} = \left[\sum_{s \in \Phi_{m,n}} \frac{\alpha_s}{\sqrt{s_1! s_2! \hdots s_m!}} (a_1^{\dag})^{s_1} (a_2^{\dag})^{s_2} \hdots (a_m^{\dag})^{s_m} \right] \ket{00\hdots0}. \]

\subsubsection{Linear optics.} The action of a linear optical experiment on our system is represented by a unitary transformation on the creation operators. We have 
\[ a_j^{\dag} \rightarrow \sum_{i=1}^m u_{ij} a_i^{\dag} \]

and we write 
\[ U = \begin{pmatrix} u_{11} & u_{12} & \hdots & u_{1m} \\ \vdots & \vdots & \vdots & \vdots \\ u_{m1} & u_{m2} & \hdots & u_{mm} \end{pmatrix} \]

the $m \times m$ unitary that stores the image of the $k$-th creation operator in its $k$-th column. We denote the operation performed on the Fock space as $\hat{U}$. Applying $\hat{U}$ on $\ket{\psi}$ gives the new state

\[ \hat{U}\ket{\psi} = \left[\sum_{s \in \Phi_{m,n}} \frac{\alpha_s}{\sqrt{s_1! s_2! \hdots s_m!}} \prod_{j=1}^m \left(\sum_{i=1}^m u_{ij} a_i^{\dag}\right)^{s_j} \right] \ket{00\hdots0} \]

where the product and sum develop like a regular multivariate polynomial in the creation operators. 

\subsubsection{Polynomial formalism.} 

For simplicity we now write $x_i$ instead of $a_i^{\dag}$ and we write $P(x)$ the $m$-variable $n$-degree homogeneous multivariate polynomial such that 

\[ \ket{\psi} = P(x) \ket{00\hdots0}. \]

\

Given a state $\ket{\psi} = P(x) \ket{00\hdots0}$, applying a unitary transformation $U$ gives the new state $\ket{\psi'} = P(U^Tx) \ket{00\hdots 0}$ with $U^Tx = \begin{pmatrix} \sum_i u_{i1} x_i \\ \vdots \\ \sum_i u_{im} x_i \end{pmatrix}$.
Similarly to \cite{aaronson2011bosonsampling} we write $U[P]$ the corresponding polynomial.

\subsection{Strong simulation of linear optics: definition and framework}

The strong simulation of a quantum system is the computation of some or all output amplitudes $\alpha_s$ or probabilities $|\alpha_s|^2$. The computation can be exact or approximate up to some additive or multiplicative error. For approximating the output probabilities, see~\cite{aaronson2014generalizing,lim2025efficientclassicalLO}. In this article we will focus on exact strong simulation.

\subsubsection{Matrix permanent.}

The permanent of a $n \times n$ matrix $A$ is defined by 
\[ \Per{A} = \sum_{\sigma \in \mathcal{S}_n} \prod_{i=1}^n a_{i,\sigma_i} \]

where the sum is over the set of permutations of $[n]$. The output amplitudes of a linear optical experiment are directly related to permanents of suitable matrices \cite{scheel2004permanents,aaronson2011bosonsampling}. In an $n$-photon $m$-mode linear optical experiment with $\ket{t}$ as input, $U$ as unitary applied, the output amplitude $\alpha_s$ for some Fock state $\ket{s}$ is given by 
\[ \bra{s}\hat{U}\ket{t}= \frac{\Per{U^{r_t}_{r_s}}}{\sqrt{t_1!t_2!\hdots t_m!s_1!s_2!\hdots s_m!}} \]

where $U_{r_s}^{r_t} = U[r_s,r_t]$ is given from the rows $r_s$ of the columns $r_t$ of $U$, with possible repetitions.
For example, consider the $3$-mode unitary 
\[ U =  \begin{pmatrix} a & b & c \\ d & e & f \\ g & h & i \end{pmatrix}. \]

With the $4$-photon input $\ket{t} = \ket{1,2,1}$, the output amplitude of the state $\ket{s} = \ket{3,0,1}$ is given by the permanent of the $4 \times 4$ matrix

\[ V =  \begin{pmatrix}
    a & b & b & c \\
    a & b & b & c \\
    a & b & b & c \\
    g & h & h & i 
\end{pmatrix} \]

up to some normalization terms. In other words, $V = U^{r_t}_{r_s}$ with $r_s = [1,1,1,3]$ and $r_t = [1,2,2,3]$.

\subsubsection{Input state of our simulations.} In this article, we restrict ourselves to the case where the input is fixed to the state with one photon in the first $n$ modes~\footnote{Note that the complexity of the simulation can depend on the inputs that are considered~\cite{marshall2023simulation}.} 
\[ \ket{1_n} = \left( a_1^{\dag}a_2^{\dag} \hdots a_n^{\dag} \right) \ket{00\hdots0} = I_{[n]}(x) \ket{00 \hdots 0} \]

where $I_{[n]}(x) = x_1x_2 \hdots x_n$. More generally we write $I_K = x_{K_1}x_{K_2} \hdots x_{K_n}$ with $K \in [m]^n$. From now on we set $P = U[I_{[n]}]$. For convenience we also write $P_j = U[x_j]$ such that 

\begin{equation}\label{eq:defP}
     P(x) = \prod_{j=1}^n \left(\sum_{i=1}^m u_{ij} x_j \right) = \prod_{j=1}^n P_j(x). 
\end{equation}

From now on, we simplify the notation by directly writing $U$ the $m \times n$ operator corresponding to the first $n$ columns of the $m \times m$ unitary that is actually applied.

\subsection{Strong simulation of linear optics: state of the art} \label{sec::soa}

\subsubsection{Exact computation with permanent formulae.} 

The state-of-the-art methods for computing one single output amplitude rely on formulae for calculating the permanent~\cite{ryser1963combinatorial,nijenhuis1978combinatorial,glynn2010permanent,tichy2011phd,shchesnovich2013asymptotic,shchesnovich2020classical}. For instance Glynn's formula~\cite{glynn2010permanent}, further improved in~\cite{tichy2011phd,shchesnovich2013asymptotic}, exploits the properties of the roots of unity. Writing $R_t$ the set of the $t$-th roots of unity, we have:

\[ \Per{U_{r_s}} = \frac{\prod_{j=1}^{m}s_j!}{\prod_{j=1}^{m}(s_j+1)} \sum_{r_1 \in R_{s_1+1}} \hdots \sum_{r_m \in R_{s_m+1}} (\overline{r_1})^{s_1}  \hdots (\overline{r_m})^{s_m} P(r_1, \hdots, r_m) \]

This formula requires the evaluation of $P$ at $\prod_{j=1}^m (s_j+1)$ different points, for a total of 

    \[ O\left(\prod_{j=1}^m (s_j+1) \times mn \right) \]

complex operations. With some extra optimizations~\cite{shchesnovich2020classical} this count can be reduced to 

    \[ O\left(\frac{\prod_{j=1}^m (s_j+1)}{\min_{s_l \neq 0} s_l+1} \times n\right). \]

To compute all amplitudes one can naively compute each permanent independently. The cost of this approach depends on the quantity 
\[ \sum_{\ket{s} \in \Phi_{m,n}} \frac{\prod_{j=1}^m (s_j+1)}{\min_{s_l\neq 0} s_l + 1} \]

for which there is no known analytical formula. However, it is known that 
\[ \sum_{\ket{s} \in \Phi_{m,n}} \prod_{j=1}^m (s_j+1) = \binom{2m+n-1}{n}  \]

(see, e.g, \cite{clifford2024faster} for a proof). We decided to ignore the $\min_{s_l\neq 0} (s_l + 1)$ term such that we approximate the cost of computing all output amplitudes as 

\[ O\left(\binom{2m+n-1}{n} \times n \right). \]

These formulae have no cost in memory other than storing the matrix and an array of $n$ coefficients.

\subsubsection{Exact computation of all the amplitudes with the SLOS algorithm.}

To our knowledge, to compute all amplitudes at once the best method is the algorithm SLOS detailed in \cite{heurtel2023slos}. It develops the product term by term. Having developed the first $k$ $P_j$'s we get a partial polynomial $P_{[k]}$ with $\binom{m+k-1}{k}$ elements and
    \[ P = P_{[k]} \times \prod_{j=k+1}^{n} P_j. \]

Developing the next term requires $m \times \binom{m+k-1}{k}$ complex multiplications and the same amount of complex additions. The total number of complex operations is 
    \[ 2\sum_{k=0}^{n-1} m \times \binom{m+k-1}{k} = 2n \binom{n+m-1}{n}. \]

Without dynamical reallocation, we need to store at least all coefficients of $P_{[k]}$ and $P_{[k+1]}$ at any step $k$. When $k=n$, we need a maximum of $\binom{m+n-1}{n} + \binom{m+n-2}{n-1}$ elements in memory. \\  

\subsubsection{SLOS with mask.} SLOS also offers the possibility to compute all the amplitudes that match a given Fock state on a subset of modes. Such Fock state is called a mask. When developing the polynomial, SLOS only needs to compute the amplitudes of intermediate states that can lead to a state that matches the mask. For instance, with the mask $x_2^2$, any term with $x_2^3$ can be ignored, which saves some computations. 

Hence, it is possible to recover all possible output amplitudes of an experiment by iterating over all possible masks on a given subset of modes and calling SLOS with mask. The memory and time complexities of such a process are derived in Appendix~\ref{appendix:slos_mask}. The time complexity only depends on the number of modes $k$ of the mask and it is given by 
\begin{equation} \sum_{s=1}^n \sum_{d=s}^{n} 2s \times \binom{n-d+k-\alpha}{n-d} \times \binom{m}{s} \times \binom{d-1}{d-s}. \label{eq::slos_mask} \end{equation}
where $\alpha$ is $1$ if $k=m$ and $0$ otherwise.

\subsubsection{Limitations and contributions.} For strong simulation, the two approaches in the state of the art both face limitations: 
\begin{itemize}
    \item[$\bullet$] developing and storing everything as done in SLOS~\cite{heurtel2023slos} is highly time-efficient but requires substantial memory, causing the memory limit to be reached quickly in practice. Once this limit is reached, the simulation can no longer run without incurring costly data movements. As a result, the computational time becomes dominated by data transfer costs, preventing us from fully leveraging the initial time efficiency.
    \item[$\bullet$] formulae for computing one coefficient at a time could be used to iterate over the set of amplitudes without encountering memory issues. However, this approach is impractical in terms of computational time, preventing us from fully benefiting from the memory savings.
\end{itemize}

We miss an intermediate method that offers good memory performance while not sacrificing the computational time. Our goal is to be able to simulate system sizes that are currently out of reach, either because memory is missing or because the computational time explodes for traditional permanent-based methods. Even though SLOS with masks can be used to obtain first trade-offs, we seek at substantially improving them, by proposing a method that only needs $O(2^n)$ in memory. As no small closed-form expression has been found, we provide the formula for the time complexity in Section~\ref{sec::slos_tree}. We detail how our method naturally extends to the simulation of adaptive linear optics and noisy simulation in Section~\ref{sec::extensions}.

\

\section{A memory-time trade-off method for exact strong simulation} \label{sec::slos_tree}

The main steps and formulas of the LO-SLAP algorithm are described in Section~\ref{subsec:overview}. The time and memory complexities are analyzed and compared to existing methods in Section~\ref{subsec:complexity}. Pseudocode and implementation details are provided in Section~\ref{sec::implem}. Finally, a method for further improving the time complexity is explored in Section~\ref{subsec:steinertree}.

\subsection{Overview of the LO-SLAP algorithm}\label{subsec:overview}

\begin{figure}[p]
    \hspace*{-1cm}
    \includegraphics{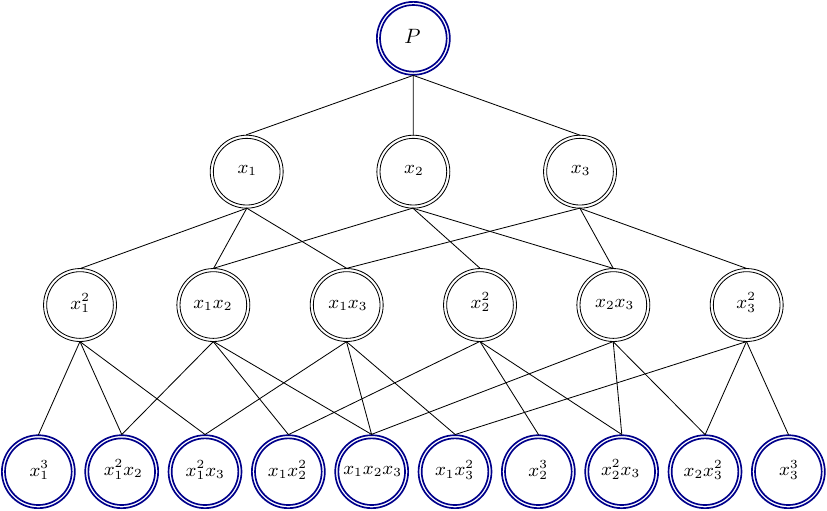}
    \caption{Example of the semi-lattice structure considered in our algorithm for $3$ photons and $3$ modes.}
    \label{fig:lattice}
    \hspace*{-1cm}
    \includegraphics{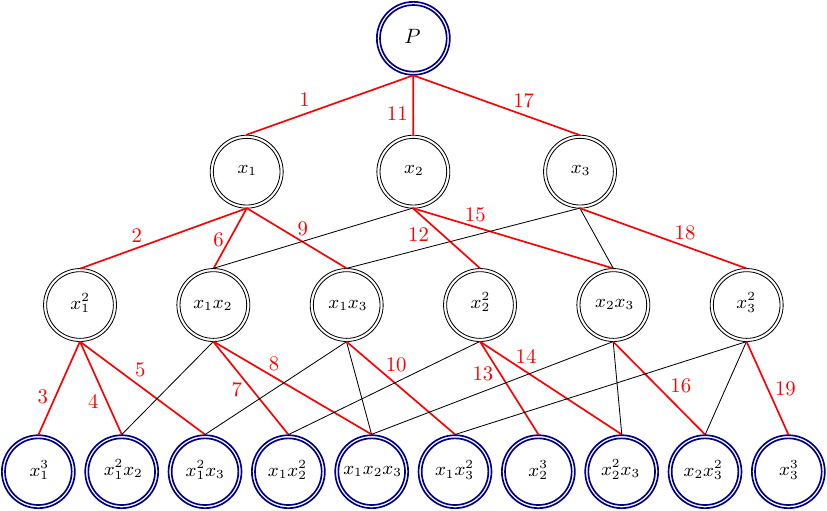}
    \caption{Illustration of the depth-first search traversal. Each red edge is labeled by its order in the traversal. Non-red edges are not used during the computation.
    }
    \label{fig:dfs}
\end{figure}

We consider a semi-lattice structure as in Fig.~\ref{fig:lattice}. Each node is labeled by a monomial $x^s = x_1^{s_1} \hdots x_m^{s_m}$ 
and contains the partial derivative\footnote{Note that formalization with derivatives can be used in generic settings, see~\cite{riemann2024}.} 
\[ \frac{\partial P}{\partial x^s}. \] 
The root is the factorized version of $P$, while the leaves, i.e. nodes of the last layer of the semi-lattice, correspond to the amplitudes we want to compute up to some known normalization coefficient. Each node has $m$ children, given from it by differentiating the associated polynomial by an $x_i, i=1 \hdots m$. A level $k$ of the lattice corresponds to all nodes of partial derivatives of same degree $k$. We will interchangeably label one node either by its monomial $x^s$ or the associated Fock state $\ket{s}$.

\

Note that the output amplitudes correspond to the leaves, i.e.\ the last layer of the lattice. Our algorithm performs a depth-first search on the lattice. When a leaf is reached, we extract the coefficient from the data of the node. This leads to a natural way to iterate over all output amplitudes of the linear optical experiment. The computation of one node is cheaper if we already know one of its parents. This provides a computational advantage over the permanent-based method, which can be seen as computing one leaf at a time without using the other nodes of the lattice. In terms of memory, we need to store at most one entire path in the lattice, as detailed later.

\

We first explicit what needs to be stored in one node and how we can efficiently compute the children of a node. Then, we analyze the global cost of traversing the lattice to visit all leaves and iterate over all output coefficients.

\

\subsubsection{The data of one node.} Given any monomial $x^s$ of degree $k \leq n$, we have 

\begin{equation} \frac{\partial P}{\partial x^s}(x) =  \sum_{J \in \mathcal{C}^n_{n-k}} \Per{U_{r_s}^{[n] \backslash J}} \prod_{j \in J} P_j(x) 
\label{eq1}
\end{equation}

where $\mathcal{C}^{n}_{n-k}$ is the set of combinations of $n-k$ elements in $[n]$.

\

One node $x^s$ at level $k$ of the lattice is therefore completely characterized by the set of $\binom{n}{k}$ coefficients 

\[ C_s = \left\{ \Per{U_{r_s}^{[n] \backslash J}}, \; J \in \mathcal{C}_{n-k}^n \right\}. \]

\

\subsubsection{Efficient computation of a child node from a parent node.} For some $k<n$, given a node $x^s, |s| = k$ and $x^{s'}, |s'|=k+1$ one of its children, we can efficiently compute $C_{s'}$ from $C_s$. Let us assume that $\ket{s}$ and $\ket{s'}$ differ in the entry $i$, we have
\[ \frac{\partial P}{\partial x^{s'}} = \frac{\partial P}{\partial x_i} \frac{\partial P}{\partial x^{s}} = \sum_{J \in \mathcal{C}^n_{n-k}} \Per{U_{r_s}^{[n] \backslash J}} \frac{\partial}{\partial x_i} \prod_{j \in J} P_j(x).  \]

Differentiating the product gives 
\[ \frac{\partial P}{\partial x^{s'}} = \sum_{J \in \mathcal{C}^n_{n-k}} \Per{U_{r_s}^{[n] \backslash J}} \sum_{\ell \in J} \frac{\partial P_\ell}{\partial x_i} \prod_{j \in J, j \neq \ell} P_j(x) \]

As $\frac{\partial P_\ell}{\partial x_i}=u_{i\ell}$ (cf~Equation~\ref{eq:defP}), we have
\begin{equation}\label{eq:node} \frac{\partial P}{\partial x^{s'}} = \sum_{J \in \mathcal{C}^n_{n-k}} \sum_{\ell \in J} \; \Per{U_{r_s}^{[n] \backslash J}} \cdot  u_{i\ell} \prod_{j \in J, j \neq \ell} P_j(x). \end{equation}

Note that by equivalently summing over $(J\in \mathcal{C}^n_{n-k-1},\ell\in[n]\backslash J)$ instead of $J\in \mathcal{C}^n_{n-k}$, we can prove Eq.~\ref{eq1} by induction as follows.  
\begin{align*}
\frac{\partial P}{\partial x^{s'}} & = \sum_{J \in \mathcal{C}^n_{n-k-1}} \left( \sum_{\ell \in [n] \backslash J} \;\; \Per{U_{r_s}^{[n]\backslash J \backslash \ell}} \cdot u_{i\ell} \right) \prod_{j \in J} P_j(x) \\
& = \sum_{J \in \mathcal{C}^n_{n-k-1}} \Per{U_{r_{s'}}^{[n] \backslash J}} \prod_{j \in J} P_j(x)
\end{align*}
where the last equality holds by the Laplace expansion along the $i$-th row of $U$ with columns $[n] \backslash J$ (see \cite{marcus1965permanents} for example).

Following Eq.~\ref{eq:node}, as the $P_j$ don't intervene, we only need to compute all the products

\[ \Per{U_{r_s}^{[n] \backslash J}} \cdot u_{i\ell} \]

for $J \in \mathcal{C}^n_{n-k}$ and $\ell \in J$. Then we accumulate the products in the corresponding element of $C_{s'}$, as summarized in Algorithm~\ref{algo:update}. The index of the sets $C_s$ and $C_{s'}$ are detailed in Section~\ref{sec::implem}. Therefore we need 
\begin{equation}\label{eq:nodecomplexity}
    |J| \times |\mathcal{C}^n_{n-k}| = (n-k) \times \binom{n}{n-k}
\end{equation}  

complex multiplications and additions to compute $C_{s'}$. 

\begin{algorithm}[H]
\caption{Update coefficient after differentiating by $x_i$ at level $k$\label{algo:update}}
\begin{algorithmic}[1]
\Procedure{update\_coefficients}{U, $x_i$, $k$, $C_{s}$, $C_{s'}$}
\For{$J \in \mathcal{C}^{n}_{n-k}$}
\State{$c \leftarrow$ $C_s[J]$}
\For{$l \in J$}
\State{prod $\leftarrow c \times U[i,l]$}
\State{$C_{s'}[J \backslash l] \leftarrow C_{s'}[J \backslash l] + \text{prod}$}
\EndFor
\EndFor
\EndProcedure
\end{algorithmic}
\end{algorithm}

\subsubsection{A depth-first search traversing algorithm.} 
While, we perform a depth-first search to attain the compute the leaves, given that we have computed of one their parent. We illustrate the behavior of the depth-first search on the example of Fig.~\ref{fig:lattice} in Fig.~\ref{fig:dfs}. The structure and the implementation are detailed in Section~\ref{sec::implem}.

\subsection{Complexity analysis and benchmarks}\label{subsec:complexity}

All the complexities are summarized in Table~\ref{tab:recap}.

\begin{table}[]

    \centering
    \renewcommand{\arraystretch}{1.3}
    \begin{tabular}{ccccccccc}
        \toprule
        \toprule
        & \hspace*{0.5cm} & Memory & \hspace*{0.5cm}& \multicolumn{5}{c}{Time} \\ 
        \cmidrule{3-3} \cmidrule{5-9}
        && && gen. & \hspace*{0.3cm} & $n=m$ & \hspace*{0.3cm} & $m \gg n$ \\
        \\
         Permanent-based  && \multirow{2}{*}{\textcolor{blue}{$m^2$}} && \multirow{2}{*}{$n \binom{2m+n-1}{n}$} &&  \multirow{2}{*}{$n \frac{6.76^n}{\sqrt{\pi n}}$} && \multirow{2}{*}{$n2^nm^{n}$} \\
         \cite{ryser1963combinatorial,glynn2010permanent,tichy2011phd,shchesnovich2013asymptotic} & && & \\ \\
         SLOS \cite{heurtel2023slos} && $\binom{n+m-1}{n}$ && \textcolor{blue}{$n \binom{n+m-1}{n}$} && $ \textcolor{blue}{n \frac{4^n}{\sqrt{\pi n}}} $ && \textcolor{blue}{$nm^{n-1}$} \\ \\
         LO-SLAP && \textcolor{purple}{$2^n$} && \textcolor{purple}{See Eq.~\ref{eq:complex}} && \textcolor{purple}{$ n \frac{5.83^n}{\sqrt{\pi n}}$} && \textcolor{blue}{$nm^{n-1}$} \\
         \bottomrule
    \end{tabular} \vspace{0.3cm}
    \caption{\label{tab:recap}Summary of the time and memory complexities of our method versus the state of the art. All complexities are $O()$ although not displayed for clarity. Best results are in blue, second best results are in purple. 
}
    
\end{table}

\subsubsection{Complexity analysis of LO-SLAP}
Let us consider the full lattice as illustrated in~Fig.~\ref{fig:lattice}. The root is at level $0$ of the lattice while the leaves are at level $n$. As described with~Eq.~\ref{eq:nodecomplexity}, a node at level $k$ of the lattice needs 

\[ 2(n-k+1) \times \binom{n}{n-k+1} = 2n \times \binom{n-1}{k-1}  \]

complex operations to be computed. Level $k$ contains all $k$-photon $m$-mode Fock states. Therefore the total number of complex operations of one level is 
\[ 2n \times \binom{n-1}{k-1} \times \binom{m+k-1}{k}.  \]

Summing over all possible levels give a total cost of 
\begin{equation} 2n\times \sum_{k=1}^n  \binom{n-1}{k-1} \times \binom{m+k-1}{m-1} \label{eq:complex} \end{equation}

complex operations. To our knowledge, no known analytical formula exists for this sum, but in the case where $n=m$ we have 
\[ 2n\times \sum_{k=1}^n  \binom{n-1}{k-1} \times \binom{n+k-1}{n-1} \sim 2n \frac{(3+2\sqrt{2})^n}{2^{5/4} \sqrt{\pi n}} \approx O\left(n \frac{5.83^n}{\sqrt{\pi n}}\right). \]

When $m \gg n$, which happens in some boson sampling settings, we can approximate the complexity by 

\[ 2n \times \sum_{k=1}^n \binom{n-1}{k-1} \times m^k \approx O(nm^{n-1}). \]

In terms of memory, we need to store at most one node per level, for a total of 
\[ \sum_{k=0}^n \binom{n}{k} = 2^n \]

complex numbers to store. 

\subsubsection{Comparison with the state of the art.}
By way of comparison, when $n=m$ the asymptotic complexity of SLOS is

\[ O\left(n \frac{4^n}{\sqrt{\pi n}}\right) \]

and when $m \gg n$ the cost is

\[ O(nm^{n-1}). \]

The permanent-based approach, when $n=m$~\cite{clifford2024faster} has a complexity of 
\[ 1.69^n \times \binom{n+m-1}{n} \sim 1.69^n \times \frac{4^n}{\sqrt{\pi n}} \approx O\left(\frac{6.76^n}{\sqrt{\pi n}} \right) \]

and when $m \gg n$ the cost is 

\[ n2^n m^{n-1}.  \]

\

\subsubsection{Benchmarks.} Overall, our method offers a clear computational advantage over permanent-based formula and a clear memory advantage over SLOS. Furthermore, as the number of modes increases for a fixed number of photons, our method becomes more and more competitive in terms of computational time compared to SLOS while not increasing the memory cost. 

\begin{figure}[h!]
    \centering
\includegraphics[width=0.9\linewidth]{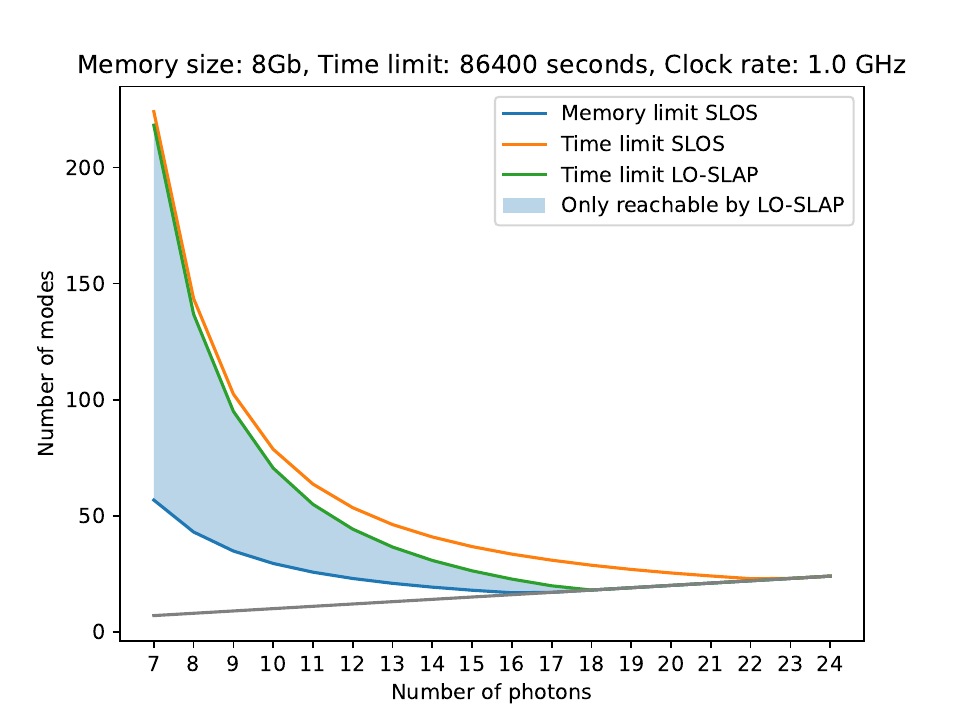}
    \caption{\label{fig::graph_slap_slos}For a given number of photons, we give the maximum number of modes we can strongly simulate with a classical computer of $8$ Gb of memory and a clock rate of $1$ GHz. The memory limit of LO-SLAP is not shown as it is the vertical line at $29$ photons.}
    
\end{figure}

\begin{figure}[h!]
\centering
\includegraphics[width=0.9\linewidth]{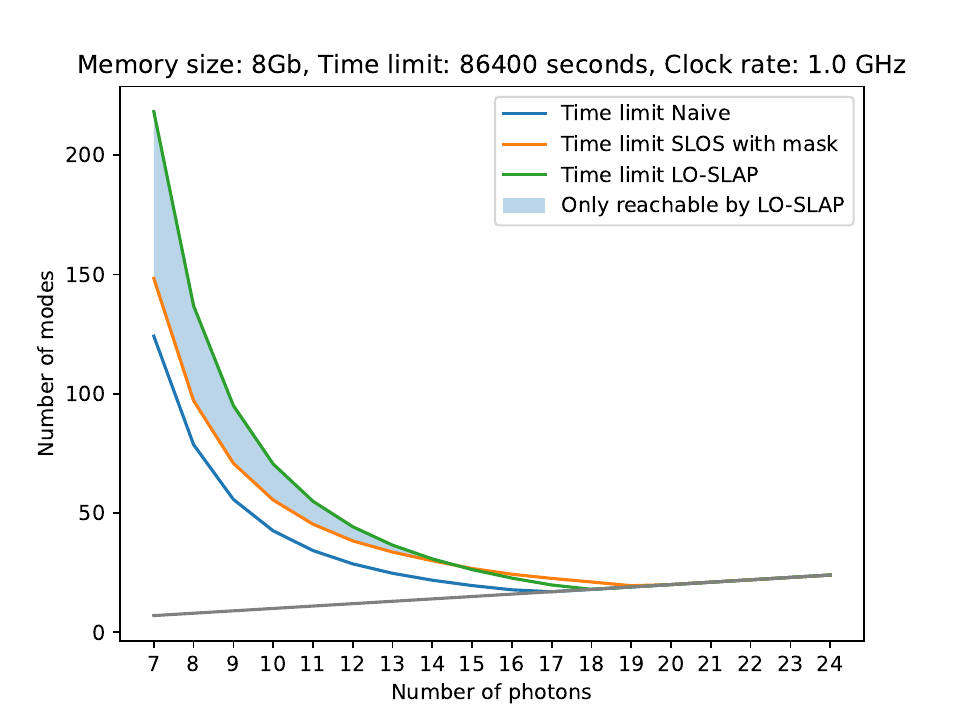}
    \caption{\label{fig::graph_slap_slos_mask} Similarly to Fig.~\ref{fig::graph_slap_slos}, we give the maximum number of modes we can strongly simulate with a classical computer of $8$ Gb of memory and a clock rate of $1$ GHz. We compare LO-SLAP against SLOS with mask and the permanent-based method. For SLOS, we took the best mask optimizing the time given the memory limit. Therefore, in this case, the fact that LO-SLAP can reach larger problem sizes is because LO-SLAP is faster than the other methods (including SLOS with mask). } 
\end{figure}

We illustrate the computational advantage of LO-SLAP with the following experiment: given a computer with $8$ Gb of memory and a clock rate at $1$ GHz, we estimate the maximum number of modes our computer can strongly simulate in a day ($86400$ seconds) as a function of the number of photons. The results are given in Fig.~\ref{fig::graph_slap_slos} for a comparison against SLOS and in Fig.~\ref{fig::graph_slap_slos_mask} for a comparison against SLOS with mask and the permanent-based method, consisting of computing every probability independently. 


\subsection{Implementation details} \label{sec::implem}

In the depth-first search of LO-SLAP, the current node is represented by a stack, where each entry refers to one variable $x_i$, with possible repetition. Therefore the size of the stack gives the degree of the current monomial. The global structure of the algorithm is given in Algorithm~\ref{algo:main}. 

\begin{algorithm}[h!]
\caption{Core algorithm: LO-SLAP main loop and subroutines\label{algo:main}}
\begin{algorithmic}[1]
\Procedure{main}{U} 
\State{flag $\leftarrow$ 1}
\State{stack $\leftarrow$ []} 
\State{\{C$_s$\}$_s \leftarrow []$}
\State{next\_element\_stack $\leftarrow$ 1}

\While{flag $> 0$}
\State{flag $\leftarrow$ compute\_next\_node(U, stack, next\_element\_stack, \{C$_s$\}$_s$)}
\If{flag $= 2$}
\State{s $\leftarrow$ get\_state\_from\_stack(stack)}
\State{C$_s$[$0$] $\leftarrow$ C$_s$[$0$]/$\sqrt{s_1!\hdots s_m!}$}
\State{yield C$_s$[$0$]}
\EndIf
\EndWhile
\EndProcedure
\\ 
\Procedure{get\_state\_from\_stack}{stack} 
  \State{s $\leftarrow$ [0]*m}
  \For{$i = 1 \hdots m$}
  \State{s[stack[i]] += 1}
  \EndFor\\
  \Return s
\EndProcedure
\\ 
\Procedure{compute\_next\_node}{U, stack, next\_element\_stack, $\{C_s\}_s$}
  \If{next\_element\_stack $> m$ \textbf{or} $|$stack$|=n$}
  \If{$|$stack$|=0$}
    \Return 0
  \Else
    \State{a $\leftarrow$ stack.pop()}
    \State{next\_element\_stack $\leftarrow$ a+1}\ \ \  \texttt{/* modified in place */}
  \EndIf
  \Else
  \State{s $\leftarrow$ get\_state\_from\_stack(stack)}
  \State{stack.add(next\_element\_stack)}
  \State{s' $\leftarrow$ get\_state\_from\_stack(stack)}
  \State{update\_coefficients(U, $x_{stack[-1]}$, $|$stack$|-1$, $C_s$, $C_{s'}$)}

  \If{feed\_forward} 
    U $\leftarrow$ update\_U(stack)
  \EndIf
  
  \EndIf
  \If{$|$stack$|$ = n}
  \Return 2
  \Else~\Return 1
  \EndIf
\EndProcedure
\end{algorithmic}
\end{algorithm}

\subsubsection{Implementation of Algorithm~\ref{algo:update}.} The main bottleneck of the algorithm is the update of the coefficients, that was informally described in Section~\ref{subsec:overview}. We recall that the update part consists in: 
\begin{enumerate}
    \item iterating over all coefficients of the parent node, 
    \item iterate over some entries of the $j$-th line of $U$,
    \item multiply the current coefficient and the current entry of $U$, \item add the result to one coefficient of the child node.
\end{enumerate}

\begin{algorithm}[h!]
\caption{Implementation of Algorithm~\ref{algo:update}}
\label{algo:implem}
\begin{algorithmic}[1]
\Procedure{update\_coefficients}{U, $x_i$, $k$, $v$}
\State{$j \leftarrow 2^{n-k}-1$}
\State{$\text{lim} \leftarrow 2^n-1$}
\While{$-1 < j < \text{lim} + 1$}
\State{$v[j] \leftarrow 0$}
\State{$j_2 \leftarrow j \oplus \text{lim}$}\ \ \ \texttt{/* $\oplus$ stands for the bitwise XOR */} 
\While{$j_2$}
\State{$p \leftarrow \_\_builtin\_ctz(j_2)$}
\State{$j_2 \leftarrow j_2 \oplus 2^p$}
\State{$v[j] \leftarrow U[i,p] \times v[j \oplus 2^p]$}
\EndWhile
\State{$t \leftarrow j \; | \; (j-1)$}\ \ \  \texttt{/* $|$ stands for the bitwise OR */}
\State{$j \leftarrow (t+1) \; | \; (((\sim t \; \& \; -\sim t) - 1) >> (\_\_builtin\_ctz(j)+1))$}
\EndWhile
\EndProcedure
\end{algorithmic}
\end{algorithm}

Fortunately we can efficiently store all the $2^n$ coefficients required in one vector of size $2^n$. There is a natural mapping between the binary representation of the indices of the vector and the coefficients. Any $\Per{U_{r_s}^{[n]\backslash J}}$ can be associated to the $n$-bit integer $b$ such that the $i$-th bit of $b$ is $1$ if $i$ is in $J$. In other words, given $v$ our vector of coefficients, $v[-1] = 1$, $v[0]$ contains the coefficients of the leaves and for instance\footnote{The binary representation of $13$ is $1101$.} $v[13]$ is the coefficient in front of the product $P_1P_3P_4$.

Given a parent node at level $k$ and $x_\ell$ the variable we want to differentiate the node with, the update part can be rephrased as: 
\begin{enumerate}
    \item iterate over all integers $i$ of Hamming weight $n-k$,
    \item iterate over the indices $j$ of the nonzero bits of $i$,
    \item set $i_2$ equal to $i$ with the $j$-th bit set to $0$,
    \item update $v[i_2] \mathrel{+}= v[i] \times U[\ell,j]$.
\end{enumerate}

This method can be compactly and efficiently implemented using bit twiddling hacks \cite{anderson2005bit}. We give the pseudo-code in Algorithm~\ref{algo:implem}. Most operations can be implemented in many different programming languages, except maybe the built in function ctz which is available in GCC and gives the number of trailing zeros in an integer starting from the least significant bit.

\subsection{Traversal time optimization with Steiner trees}
\label{subsec:steinertree}
\begin{figure}[p]
    \hspace*{-1cm}
    \centering\includegraphics{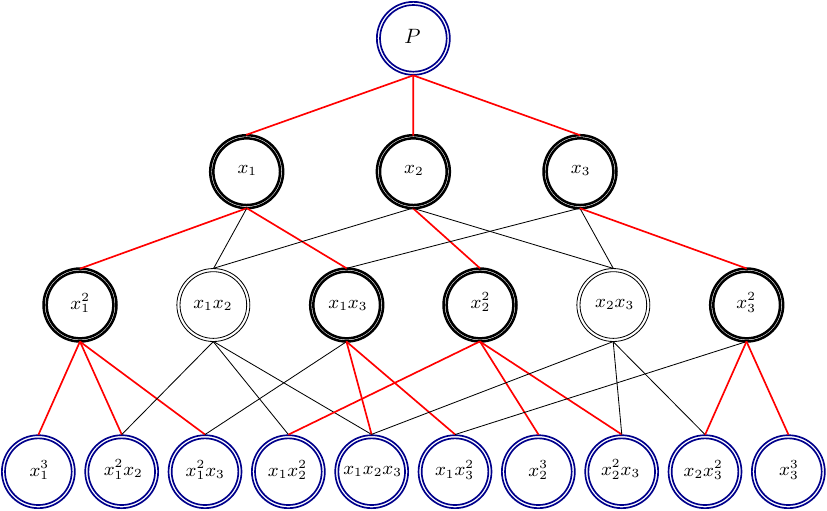}
    \caption{Optimized traversal of the graph. The nodes $x_1x_2$ and $x_2x_3$ are not visited anymore which will reduce the computational complexity.}
    \label{fig:compact}
    \hspace*{-0.5cm}
    \centering\includegraphics[width=\textwidth]{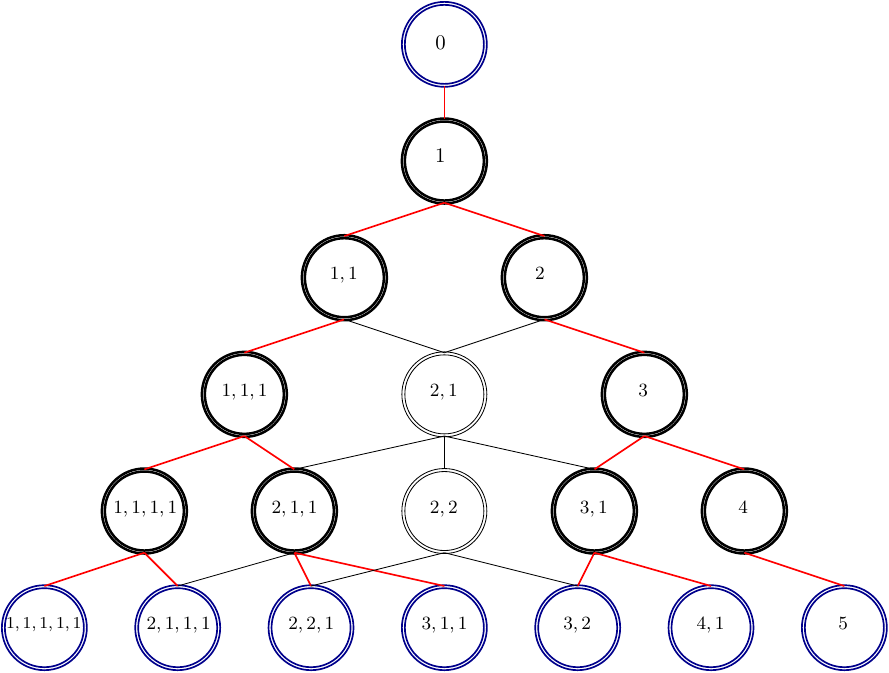}
    \caption{Reduced version of a lattice with $5$ photons. Each node is now an equivalence class. The two nodes $(2,2)$ and $(2,1)$ not in bold are two nodes that will not be visited in an optimal traversal of the lattice with $5$ photons. Note however that for $4$ photons we need to visit these two nodes.}
    \label{fig:compress}
\end{figure}

The naive depth-first search we use for the traversal of the lattice does not exploit the high connectivity of the lattice. If the goal is only to reach the leaves, more efficient traversals can be designed to avoid redundant nodes and save computational operations. For instance, still in our example of Fig~\ref{fig:lattice}, we can skip the nodes $x_1x_2$ and $x_2x_3$ but still visit all the leaves, as described in Fig~\ref{fig:compact}. Finding a more efficient traversal would be very useful to reduce even further the total computational time.

The problem of finding the optimal traversal is closely related to the Steiner tree problem~\cite{hwang1992steinertree}. In its usual formulation, we are given a graph $G = (V,E)$ with weights on the edges and a set of vertices $S$ called terminals. A Steiner tree is a tree in $G$ that spans $S$; the Steiner tree problem consists in finding the minimum weight Steiner tree, where the weight is given by the sum of the weights of the edges. Variants include the directed case, also called the directed Steiner tree problem or the Steiner arborescence problem. In this case we are given a root $r$, a set of terminals $S$ and we look for an arborescence in $G$ rooted at $r$ spanning $S$.

The lattice we consider has a natural interpretation as a directed graph, where each node is connected to its parents by a directed edge from the parent to the node. The weight of an edge is the number of complex operations required to compute the coefficients of the tail of the edge. Setting the root to the top node of the lattice and the terminals to the leaves, the solution of the associated directed Steiner tree problem would give the optimal traversal of the lattice.

The generic Steiner tree problem is known to be NP-hard \cite{garey1977complexitysteinertree}. We do not know if optimal solutions can be found in polynomial time on the specific graph structure of the lattice. To find a Steiner tree and provide numerical evidence that it could indeed improve traversal time, we use a heuristic in the SCIP-Jack package \cite{RehfeldtKoch2023} which achieved top rankings at PACE 2018 \cite{pace2018}.

\subsubsection{Reducing the graph size.} The main issue with this approach so far is the size of the graph. With $n$ photons and $m$ modes the total number of nodes in the lattice is 
\[ \sum_{k=0}^n \binom{k+m-1}{k} = \binom{n+m}{n}. \]

This is larger than the memory requirement of the SLOS algorithm \cite{heurtel2023slos} which is theoretically much faster than our method. Not only we lose the memory advantage, but most likely the computation of the Steiner tree will be very costly as well, even though we only have to execute it once for fixed $n$ and $m$. Still, it seems unlikely that computing a Steiner tree would provide any advantage. 

\

\subsubsection{Reduced lattice.} Instead of applying the Steiner tree problem on the full lattice, we propose to apply it on a reduced version. All the monomials that are equal up to a change of variables are gathered in one node. We illustrate it in Fig~\ref{fig:compress} with an example on $5$ photons. For example, $x_1^3$, $x_2^3$ and $x_3^3$ are all merged into one node labeled $(3)$. Similarly, $x_1x_2$, $x_1x_3$ and $x_2x_3$ are merged into a node of label $(1,1)$. Therefore, each node is a canonical representative of an equivalence class of nodes in the original lattice. The new graph preserves the structure of the lattice: two nodes in the original lattice are connected only if their canonical representatives are connected in the new graph. 

At level $k$, each node is encoded by a different partition of the integer $k$. The number of nodes at a level $k$ is therefore given by the number of partitions $p(k)$ of the integer $k$. The total number of nodes in this new graph is 
\[ \sum_{k=0}^n p(k). \]

Even for $n = 40$, which is way beyond what we could achieve, the total number of nodes is $215308$. For comparison, the original lattice with $n=m=40$ contains approximately $5 \times 10^{22}$ nodes. Note also that the size of our new graph no longer depends on the number of modes.

All the nodes in an equivalence class have the same computational cost. The weight of an edge is the cost of one node multiplied by the number of nodes in the equivalence class.

\subsubsection{Details on an efficient implementation.} 

Given a solution to the Steiner tree problem on this reduced lattice, we need to design an efficient way to perform the traversal. Indeed, when visiting an equivalence class of the reduced lattice, we want to avoid visiting the same node multiple times. Furthermore, checking if a node has been visited could lead to significant time or memory overhead.

Instead of storing a stack to represent the current state, we will use a vector $v$ of size $n+1$ where each entry $v[i]$ contains the set of variables $(x_j)_j$ whose degree is $i$ in the monomial. For instance, for the $8$-photon Fock state $\ket{s} = \ket{2,0,1,2,0,3}$ we have 

\[ v = [\{x_2, x_5\}, \{x_3\}, \{x_1, x_4\}, \{x_6\}, \emptyset, \emptyset, \emptyset, \emptyset, \emptyset]. \]

Each $v$ is associated to one node in the reduced graph with the partition of $n$

\[ n = \sum_{i=0}^{n} |v[i]| \times i. \]

Computing the children of a node in the original lattice now consists in choosing one variable from a set $v[i]$ of $v$ and moving it to the set $v[i+1]$. To respect the traversal given by the Steiner tree, we look at the children of the equivalent node in the reduced graph. This will give us which entries $i$ of $v$ have to be modified. For each such entry $i$ we iterate over the set $v[i]$, move one variable and we perform our computations just like in the regular case. At this point this is strictly equivalent to adding a variable in our stack.

With this approach, there remains the problem of visiting the same node multiple times within an equivalent class. This can occur when an entry $v[i]$ contains more than one element. In such cases, nodes might be visited as many times as there are distinct ways to ``fill" the set $v[i]$. To avoid this, we need to impose an order ensuring that the set $v[i]$ is constructed only once. Specifically, we require that a variable $x_j$ cannot be added to $v[i]$ if there is already a variable $x_k$ with $k > j$ present.

In practice, we use hash tables to represent an entry $v[i]$, with insert and delete doable in $O(1)$. We need to store the maximum element of a set $v[i]$, which may require up to $O(n)$ to be computed when removing an element from $v[i]$. This is the most costly operation that we introduce with this approach.

\subsubsection{Theoretical gains.}

Although the size of the reduced graphs only depends on the number of photons $n$, the weights of the edges depend on the number of nodes in the original graph that are in each equivalent class. Hence the input of the Steiner tree problem depends on the number of modes $m$: an optimal solution for $m=n$ may not be an optimal solution for $m>n$. For simplicity, we only run the optimizations in the case $m=n$ and we keep the results as a basis for larger $m$ as well. 

We show in Table~\ref{tab:steiner} the theoretical gains we obtain after running the optimizations. We managed to get results up to $14$ photons. Overall, we get significant savings in the total number of FLOPs, up to $90\%$ savings, and the gains increase with the number of photons for similar ratios $m/n$. 

\begin{table}[H]
    \centering
    \caption{Summary of the theoretical gains obtained with the Steiner arborescence optimization. $n$ is the number of photons, $m$ the number of modes. In bold are shown the cases where more than $50\%$ of FLOPs gain is achieved.}
    \scalebox{0.6}{
    \begin{tabular}{cccccclccccccccc}
    \toprule
    \toprule
         & \hspace*{0.5cm} & \multicolumn{3}{c}{Nodes in partition graph} &\hspace*{0.5cm}& &&& \multicolumn{3}{c}{Nodes in original graph} &\hspace*{0.5cm}& \multicolumn{3}{c}{FLOPs}  \\
        \cmidrule{3-5} \cmidrule{10-12} \cmidrule{14-16}
         & & Original & Optim. & Gain & & & & & Original & Optim. & Gain & & Original & Optim. & Gain \\
         \\
         \multirow{3}{*}{$n=2$} & & \multirow{3}{*}{$4$} & \multirow{3}{*}{$4$} & \multirow{3}{*}{$0\%$} & & $m=2$ & & & $6$ & $6$ & $0\%$ & & $20$ & $20$ & $0\%$ \\
         & & & & & & $m=4$ & & & $15$ & $15$ & $0\%$ & & $56$ & $56$ & $0\%$\\
         & & & & & & $m=20$ & & & $231$ & $231$ & $0\%$ & & $920$ & $920$ & $0\%$ \\
         \\
         \multirow{3}{*}{$n=3$} & & \multirow{3}{*}{$7$} & \multirow{3}{*}{$7$} & \multirow{3}{*}{$0\%$} & & $m=3$ & & & $20$ & $20$ & $0\%$ & & $150$ & $150$ & $0\%$ \\
         & & & & & & $m=6$ & & & $84$ & $84$ & $0\%$ & & $624$ & $624$ & $0\%$ \\
         & & & & & & $m=30$ & & & $39711$ & $39711$ & $0\%$ & & $249240$ & $249240$ & $0\%$ \\
         \\
         \multirow{3}{*}{$n=4$} & & \multirow{3}{*}{$12$} & \multirow{3}{*}{$12$} & \multirow{3}{*}{$0\%$} & & $m=4$ & & & $70$ & $70$ & $0\%$ & & $1032$ & $1032$ & $0\%$ \\
         & & & & & & $m=8$ & & & $495$ & $495$ & $0\%$ & & $6448$ & $6448$ & $0\%$ \\
         & & & & & & $m=40$ & & & $135751$ & $135751$ & $0\%$ & & $1282800$ & $1282800$ & $0\%$ \\
         \\
         \multirow{3}{*}{$n=5$} & & \multirow{3}{*}{$19$} & \multirow{3}{*}{$17$} & \multirow{3}{*}{$10\%$} & & $m=5$ & & & $252$ & $222$ & $12\%$ & & $6810$ & $5210$ & $23\%$ \\
         & & & & & & $m=10$ & & & $3003$ & $2868$ & $4\%$ & & $64120$ & $56920$ & $11\%$ \\
         & & & & & & $m=50$ & & & $3478761$ & $3475086$ & $0\%$ & & $44715600$ & $44519600$ & $0\%$ \\
         \\
         \multirow{3}{*}{$n=6$} & & \multirow{3}{*}{$30$} & \multirow{3}{*}{$27$} & \multirow{3}{*}{$10\%$} & & $m=6$ & & & $924$ & $774$ & $16\%$ & & $43836$ & $29436$ & $33\%$ \\
         & & & & & & $m=12$ & & & $18564$ & $17112$ & $8\%$ & & $622896$ & $488256$ & $22\%$ \\
         & & & & & & $m=60$ & & & $90858768$ & $90649908$ & $0\%$ & & $1524786000$ & $1505882400$ & $1\%$ \\
         \\
         \multirow{3}{*}{$n=7$} & & \multirow{3}{*}{$45$} & \multirow{3}{*}{$36$} & \multirow{3}{*}{$20\%$} & & $m=7$ & & & $3432$ & $2585$ & $25\%$ & & $277550$ & $141428$ & $49\%$ \\
         & & & & & & $m=14$ & & & $116280$ & $99445$ & $14\%$ & & $5956664$ & $3627792$ & $39\%$ \\
         & & & & & & $m=70$ & & & $2404808340$ & $2391626465$ & $1\%$ & & $51221188760$ & $49597362080$ & $3\%$ \\
         \\
         \multirow{3}{*}{$\bm{n=8}$} & & \multirow{3}{*}{$67$} & \multirow{3}{*}{$53$} & \multirow{3}{*}{$21\%$} & & $\bm{m=8}$ & & & $\bm{12870}$ & $\bm{9034}$ & $\bm{30\%}$ & & $\bm{1736720}$ & $\bm{711248}$ & $\bm{59\%}$ \\
         & & & & & & $m=16$ & & & $735471$ & $626391$ & $15\%$ & & $56321120$ & $33916640$ & $40\%$ \\
         & & & & & & $m=80$ & & & $64276915527$ & $63987570127$ & $0\%$ & & $1702033764320$ & $1662858005600$ & $2\%$ \\
         \\
         \multirow{3}{*}{$\bm{n=9}$} & & \multirow{3}{*}{$97$} & \multirow{3}{*}{$73$} & \multirow{3}{*}{$25\%$} & & $\bm{m=9}$ & & & $\bm{48620}$ & $\bm{33518}$ & $\bm{31\%}$ & & $\bm{10771506}$ & $\bm{3557970}$ & $\bm{67\%}$ \\
         & & & & & & $\bm{m=18}$ & & & $\bm{4686825}$ & $\bm{3868275}$ & $\bm{17\%}$ & & $\bm{527992920}$ & $\bm{256217184}$ & $\bm{51\%}$ \\
         & & & & & & $m=90$ & & & $1731030945644$ & $1719603126704$ & $1\%$ & & $56092374315180$ & $53969620537620$ & $4\%$ \\
         \\
         \multirow{3}{*}{$\bm{n=10}$} & & \multirow{3}{*}{$139$} & \multirow{3}{*}{$97$} & \multirow{3}{*}{$30\%$} & & $\bm{m=10}$ & & & $\bm{184756}$ & $\bm{123781}$ & $\bm{33\%}$ & & $\bm{66348900}$ & $\bm{17957700}$ & $\bm{73\%}$ \\
         & & & & & & $m=20$ & & & $30045015$ & $25232885$ & $16\%$ & & $4916825680$ & $2498064880$ & $49\%$ \\
         & & & & & & $m=100$ & & & $46897636623981$ & $45351094447131$ & $3\%$ & & $1836655872773600$ & $1544330460713600$ & $16\%$ \\
         \\
         \multirow{3}{*}{$\bm{n=11}$} & & \multirow{3}{*}{$195$} & \multirow{3}{*}{$128$} & \multirow{3}{*}{$34\%$} & & $\bm{m=11}$ & & & $\bm{705432}$ & $\bm{463531}$ & $\bm{34\%}$ & & $\bm{406441926}$ & $\bm{86382406}$ & $\bm{79\%}$ \\
         & & & & & & $\bm{m=22}$ & & & $\bm{193536720}$ & $\bm{158971574}$ & $\bm{18\%}$ & & $\bm{45542059904}$ & $\bm{18400415836}$ & $\bm{60\%}$ \\
         & & & & & & $m=55$ & & & $1074082795968$ & $998028708078$ & $7\%$ & & $87690142188220$ & $63328604101510$ & $28\%$ \\
         \\
         \multirow{3}{*}{$\bm{n=12}$} & & \multirow{3}{*}{$272$} & \multirow{3}{*}{$176$} & \multirow{3}{*}{$35\%$} & & $\bm{m=12}$ & & & $\bm{2704156}$ & $\bm{1728918}$ & $\bm{36\%}$ & & $\bm{2478591000}$ & $\bm{429441480}$ & $\bm{83\%}$ \\
         & & & & & & $\bm{m=24}$ & & & $\bm{1251677700}$ & $\bm{1025249324}$ & $\bm{18\%}$ & & $\bm{419980534176}$ & $\bm{143902408512}$ & $\bm{66\%}$ \\
         & & & & & & $m=60$ & & & $15363284301456$ & $14884985559986$ & $3\%$ & & $1583783064511080$ & $1212435786358680$ & $23\%$ \\
         \\
         \multirow{3}{*}{$\bm{n=13}$} & & \multirow{3}{*}{$373$} & \multirow{3}{*}{$225$} & \multirow{3}{*}{$40\%$} & & $\bm{m=13}$ & & & $\bm{10400600}$ & $\bm{6813874}$ & $\bm{34\%}$ & & $\bm{15058389450}$ & $\bm{2168746606}$ & $\bm{86\%}$ \\
         & & & & & & $\bm{m=26}$ & & & $\bm{8122425444}$ & $\bm{6616339954}$ & $\bm{19\%}$ & & $\bm{3858768103216}$ & $\bm{1391546563836}$ & $\bm{64\%}$ \\
         & & & & & & $m=39$ & & & $635013559600$ & $558078517806$ & $12\%$ & & $161828613214632$ & $85970855413170$ & $47\%$ \\
         \\
         \multirow{3}{*}{$\bm{n=14}$} & & \multirow{3}{*}{$508$} & \multirow{3}{*}{$293$} & \multirow{3}{*}{$42\%$} & & $\bm{m=14}$ & & & $\bm{40116600}$ & $\bm{25664344}$ & $\bm{36\%}$ & & $\bm{91194804876}$ & $\bm{10195561460}$ & $\bm{89\%}$ \\
         & & & & & & $\bm{m=20}$ & & & $\bm{1391975640}$ & $\bm{1028365800}$ & $\bm{26\%}$ & & $\bm{1690514536480}$ & $\bm{300732164320}$ & $\bm{82\%}$ \\
         & & & & & & $\bm{m=28}$ & & & $\bm{52860229080}$ & $\bm{43509128034}$ & $\bm{18\%}$ & & $\bm{35343854172032}$ & $\bm{10686628719536}$ & $\bm{70\%}$ \\
         \\
    \bottomrule
    \end{tabular}}  
    \label{tab:steiner}
\end{table}

\section{Extensions to feedforward and noisy simulation} \label{sec::extensions}

\subsection{Feedforward}

In its usual formulation, feedforward or adaptive linear optics is the ability to apply different unitaries depending on intermediate measurements. This is usually represented with multiple layers of computation, each layer being a unitary followed by some measurements, measurements that will adaptively decide the next layer, and so on. Given $m$ modes, $n$ photons and $k$ adaptive measurements, an adaptive linear optical computation can be summarized by the data of all possible unitaries $\{ U_p \; | \; p \in \Phi_{k,r}, 0 \leq r \leq n \}$ on $m$ modes that are applied as a function of the measurement outcome $p$ on those $k$ modes.

Without loss of generality we can assume that the modes to be measured are always the first ones. 
Otherwise, we can rearrange the order by swapping the rows of $U$.

Note that as long as a set of variables $(x_i)_{i \in J \subset [n]}$ has not appeared yet in the nodes while doing the traversal, we can apply some unitary operations on the associated modes $i \in J$, only modifying the rows $i \in J$ of $U$. This will alter the output results related to the modified modes without affecting the others, making all the previous computed nodes still valid and useful. 

In practice, when a new node is computed, we need to check whether the added variable is a measured mode, and if so, we must update $U$ accordingly. This corresponds to line 36 in Algorithm~\ref{algo:main}. Note that the update of $U$ can be done simultaneously with the update of the coefficients, as the rows of $U$ that are modified do not affect the update. Overall, LO-SLAP can handle feedfoward with the only cost of updating $U$.

\subsubsection{Comparison against the state of the art.}

Even though it was not detailed in the original paper~\cite{heurtel2023slos}, SLOS with mask can also be used to strongly simulate linear optical circuits with feedforward. If $k$ is the total number of adaptive measurements, fixing a measurement outcome on those $k$ modes gives a unique matrix $U$ that is applied on the whole system. Then it is possible to use SLOS with mask to perform a restricted computation for this specific measurement outcome. Repeating such computation for any possible outcome gives a complete description of the output amplitudes. Without any particular restriction on the memory we can therefore assume that $k$, the number of adaptive measurements, is also the mask size used in SLOS. 

We recall here Eq.~\ref{eq::slos_mask} (see~Appendix~\ref{appendix:slos_mask}), the complexity of SLOS with a mask on $k$ modes: 
\[ \sum_{s=1}^n \sum_{d=s}^{n} 2s \times \binom{n-d+k-\alpha}{n-d} \times \binom{m}{s} \times \binom{d-1}{d-s}\]

where $\alpha=1$ when $k=m$ and $0$ otherwise. 

\begin{figure}[!htb]
    \centering
    \includegraphics[width=1\linewidth]{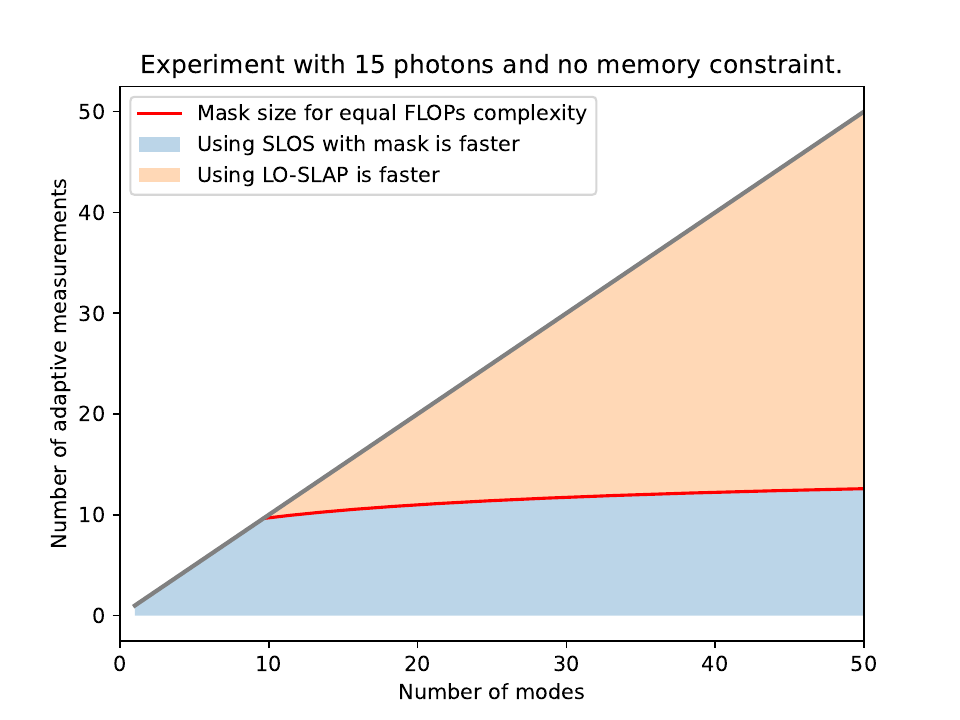}
    \caption{Regions indicating which of LO-SLAP or SLOS with mask is faster, based on the theoretical number of FLOPs, as a function of the number of adaptive measurements and the number of modes. The number of photons is fixed at $15$.}
    \label{fig::graph_slap_slos_ff}
\end{figure}

For $k=0$, i.e. with no adaptive measurements, we recover the standard version of SLOS which is the fastest method. For $k=m$, we recover the permanent-based approach which is the slowest. LO-SLAP stands between these two extreme cases and, for a given $n$ and $m$, there exists a $k$ such that SLOS with mask and LO-SLAP have the same computational complexity. This is illustrated in Fig.~\ref{fig::graph_slap_slos_ff}, where the number of photons has been fixed to an arbitrary value of $15$. The plane is divided into two regions where each method is faster. When the number of adaptive measurements is large enough, LO-SLAP prevails.

\begin{figure}[!htb]
    \centering
    \includegraphics[width=1\linewidth]{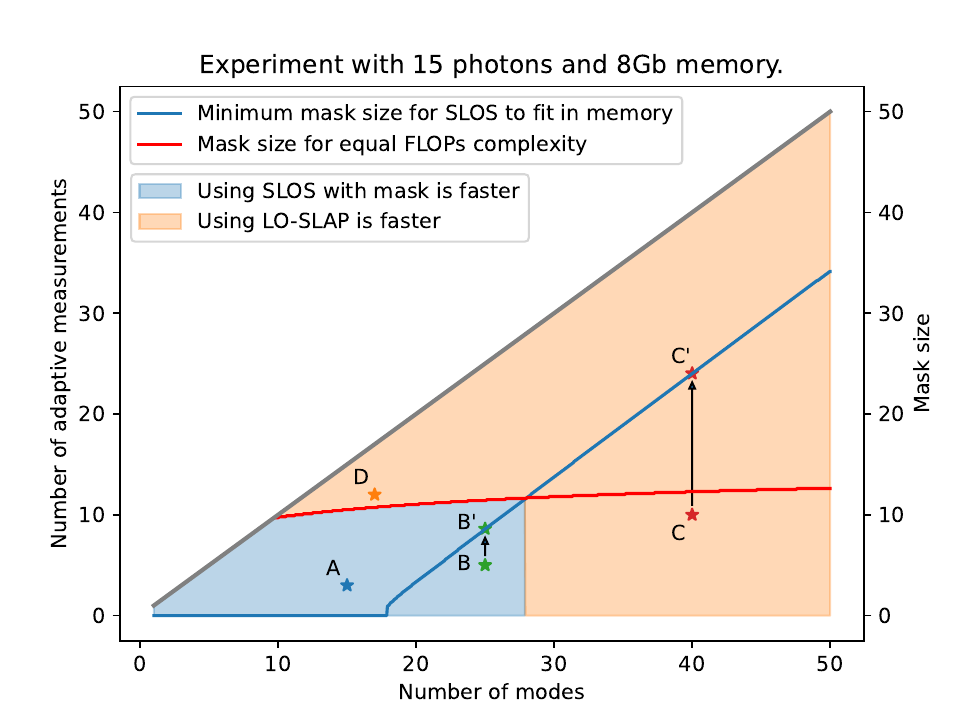}
    \caption{Same experiment than Fig~\ref{fig::graph_slap_slos_ff} with an additional memory constraint. The red curve gives the mask size for SLOS to have the same complexity than LO-SLAP. The blue curve represents the smallest possible mask for SLOS to run with 8Gb of memory --- thus, any point below the blue curve exceeds the memory limit. For instance, points A and D are above the blue curve so setting the mask size to the number of adaptive measurements is possible. Note that for point A, the mask size is smaller than the red curve so SLOS with mask is faster, whereas for point D, LO-SLAP is better. Regarding B and C, both are under the blue curve which means that the configuration is unfeasible if we keep the mask size equal to the number of adaptive measurements. We need to set the mask size to larger values, giving points B' and C'. Note that point B' is still under the red curve so that SLOS with mask remains faster. However, point C' lies above the red curve, hence LO-SLAP is faster.}
    \label{fig::graph_slap_slos_ff_bis}
\end{figure}

\subsubsection{Adding memory constraints.} So far we made no assumption on the memory to produce Fig~\ref{fig::graph_slap_slos_ff}. However on a modest laptop, even with a small $k$, if $m$ is large enough SLOS with a mask of size $k$ may not fit in the memory. If this is the case we would need a larger mask. Namely,
\[ \text{mask size} = \max(k, \text{minimum mask size for SLOS to fit in memory}) \]

and this will have an impact on which method is faster, especially for small $k$ and large $m$. We adapted Fig~\ref{fig::graph_slap_slos_ff} with a memory constraint of $8$ Gb. The results are given in Figure~\ref{fig::graph_slap_slos_ff_bis}. The number of adaptive measurements and the actual mask size used in SLOS are now two different quantities. For a given point in the plane ($m = \#$ modes, $k = \#$ adaptive measurements) we now need to check if the memory constraint can be satisfied with a mask size $k$ or if we need a larger mask. If our point $(m,k)$ is below the minimum mask size required by the memory constraint then the actual mask size needs to be adjusted, giving a new point $(m,k')$. This is illustrated with the points B and C which correspond to unfeasible configurations, and are updated to B' and C'. If the new point is below the curve giving the mask size for equal complexity between SLOS and LO-SLAP then SLOS with mask is faster, otherwise LO-SLAP is faster.

The main point we want to emphasize is that once the minimum mask size for SLOS to fit in the memory is larger than the mask size for SLOS and LO-SLAP to be of same computational complexity, then in any case LO-SLAP will be faster. Because the memory constraint imposes that the mask size will always be larger than the mask size required for SLOS and SLAP to be of similar efficiency, whatever the number of adaptive measurements.

\subsection{Noisy simulation}

We review three kinds of noise and show how LO-SLAP can handle them efficiently. Note that we only look at the exact computation of the output probabilities. For noisy weak simulation, i.e. sampling, see~\cite{garciapatron2019simulatingboson,oh2023classicalsimulation,shchesnovich2019noiseboson}.

\subsubsection{Uniform photon loss.} Photons can be lost anywhere during the computation. Different models for loss exist, here we assume that each photon has the same probability $\eta \in [0,1]$ of being lost in any mode. With such a uniform loss, we can use commutation rules for loss \cite{oszmaniec2018classical} to assume that, without loss of generality, the loss only happens at the beginning of the circuit, before applying the unitary $U$.

With this settings, the input of our system is no longer the Fock state $a_1^{\dag} a_2^{\dag} \hdots a_n^{\dag} \ket{00\hdots0}$ but a probabilistic mixture of all loss scenarios

\[ \sum_{i=0}^{n} \sum_{J \in \mathcal{C}^{n}_{n-i}} \eta^{n-i}(1-\eta)^{i} \prod_{j \in [n]\backslash J} a_j^{\dag} \ket{00\hdots0} \bra{00\hdots0} \prod_{j \in [n]\backslash J} a_j \]

where the sum on $i$ is on the number of non lost photons and the second sum is over all possible input Fock states with $i$ photons among the initial $n$ ones.

Given a state $\ket{s}$ on $k \leq n$ photons, the probability of measuring $\ket{s}$ is given by linearity over all canonical input states:
\[ p(s) = \eta^{n-k}(1-\eta)^{k} \times \sum_{J \in \mathcal{C}^n_{n-k}} \frac{\left|\Per{U_{r_s}^{[n]\backslash J}}\right|^2}{s_1!s_2! \hdots s_m!}.  \]

Fortunately, all the coefficients $\left\{ \Per{U_{r_s}^{[n]\backslash J}} \; | \; J \in \mathcal{C}_{n-k}^n \right\}$ are computed in the node associated to state $\ket{s}$ during the traversal of the LO-SLAP algorithm. See Section~\ref{sec::slos_tree} for more details. In other words, naturally, all the coefficients for any loss scenario are already computed in one call to the lossless LO-SLAP algorithm.

\subsubsection{Distinguishability.} For photons to interact they must be indistinguishable. It is equivalent to considering groups of indistinguishable photons and run the simulation on each group independently. For one group this is also completely equivalent to having lost all the photons that are not in the group. This means that for any scenario of indistinguishable groups, there is an equivalent set of loss scenarios that would give the same outputs, modulo some linear recombinations of mixtures. Therefore LO-SLAP also gives all the data necessary to compute outputs with non perfect indistinguishability. 

\subsubsection{Multi-photon emission.} At source, there is a nonzero probability that more than one photon is emitted. Therefore, we should also consider, with some probability, inputs of $n+1, n+2, \hdots,$ photons. Given that the probability that one more photon is emitted is quite small, the chances of having more than $2$ photons in each mode are quite unlikely. We can approximate the total possible number of photons to $2n$ and, based on the fact that all other scenarios are covered by our method, a single run of LO-SLAP with the input $a_1^{\dag}a_1^{\dag}a_2^{\dag}a_2^{\dag} \hdots a_n^{\dag}a_n^{\dag} \ket{00\hdots 0}$ will deal with all possible smaller number of photons as well, including both uniform loss and distinguishability.

\section{Conclusion} \label{sec::conclu}

We presented LO-SLAP, a memory efficient algorithm for the exact strong simulation of linear optical circuits. The first motivation of the algorithm was to provide better memory-time trade-offs compared to the state of the art and we showed significant improvements in the sizes of problems that we can reach with our method versus the state of the art methods. 

Our algorithm naturally extends to the simulation of feedforward and noise. In both cases, minimal extra computations are required. For the feedforward, LO-SLAP scales much more favorably with the number of adaptive measurements compared to the state of the art, while still maintaining its memory-time trade-off. Regarding noisy simulation, our algorithm demonstrates that, in practice, the strong simulation of linear optical circuits with loss and distinguishability is comparable to the simulation of noiseless circuits, up to the potential additional cost of recombining the computed coefficients. 

Our work can be further developed in several ways. From a theoretical point of view, it would be interesting to find optimal analytical solutions to the Steiner tree problem. Given that the lattice has a specific structure, we may expect that a structured solution emerges as well. More practically, incorporating multithreading or distributed computing would help in improving the performance of the algorithm. 

Overall, we expect this new algorithm to provide valuable help to any user experimenting on noisy adaptive linear optical circuits. 

\section*{Acknowledgements}
The authors would like to thank Shane Mansfield and Jean Senellart for their support and feedback. The authors also thank Daniel Rehfeldt for providing a version of SCIP-Jack and for his help. This work has been co-funded by the Horizon-CL4 program under the grant agreement 101135288 for EPIQUE project, by the ANR-24-QUA2-007-003 for ResourcesQ project and and by the CIFRE n° 2022/0081.

\bibliographystyle{splncs04}
\bibliography{mybibliography}

\appendix

\section{Complexity of SLOS with mask} \label{appendix:slos_mask}

Given an experience with $m$ modes, $n$ photons, we derive the complexity of computing all output amplitudes using SLOS with masks on $k$ modes.

\subsection{Time complexity}

We start by recalling how SLOS works and how amplitudes of intermediate states are generated. Suppose we have expanded the first $p$ terms of the polynomial: 
\[ P = P_{[p]} \times \prod_{i=p+1}^n P_i, \]

where $P_i(x) = \sum_{j=1}^m u_{ji} x_j$.

\

Instead of evaluating globally the cost of developing the next term, we need to look at the exact cost of computing one specific amplitude. Any monomial $x^{\bold{a}}$ of degree $p+1$ will be given by its "parent" monomials of degree $p$ $x^{\bold{a'}}$ where $a$ and $a'$ only differs by one unit in one entry. In other words, if we note $c_{\bold{a}}$ the amplitude of $x^{\bold{a}}$, we have 
\[ c_{\bold{a}} = \sum_{\substack{\bold{a'} \text{ parent of $\bold{a}$} \\ \text{differs in entry $j$}}} c_{\bold{a'}} u_{ji}. \]

Writing $s_{\bold{a}} = \supp{\bold{a}}$ the support of $\bold{a}$, i.e, its number of nonzero entries, then computing $c_{\bold{a}}$ needs $s_{\bold{a}}$ complex multiplications and the same amount of complex additions. Then, for a given mask $\mathfrak{m}$, the cost of a call to SLOS with this mask is given by the sum of all states that have to be generated, i.e, all states from which a state with mask $\mathfrak{m}$ can be generated. In other words,

\[ SLOS_{n,m}(\mathfrak{m}) = \sum_{\substack{\text{monomial }\bold{a} \text{ is a parent} \\ \text{of any monomial with mask $\mathfrak{m}$}}} 2s_{\bold{a}}.  \]

As explained in Section~\ref{sec::background}, for an experiment with $n$ photons, $m$ modes and a mask on $k$ modes, we need to iterate over all possible mask values on $k$ modes. The total time complexity is

\[ \sum_{\text{mask $\mathfrak{m}$ on $k$ modes}} \#SLOS_{n,m}(\mathfrak{m}) =  \sum_{\text{mask $\mathfrak{m}$ on $k$ modes}} \sum_{\substack{\text{monomial }\bold{a} \text{ is a parent} \\ \text{of any monomial with mask $\mathfrak{m}$}}} 2s_{\bold{a}}. \]

We commute the two sums to have a simpler way of computing the quantity: 
\[ \sum_{\text{monomial $\bold{a}$}} \sum_{\substack{\text{mask $\mathfrak{m}$ on} \\ \text{$k$ modes} \\ \text{reachable from $\bold{a}$}}} 2s_{\bold{a}} \]

where the first sum is over all possible Fock states on $m$ modes and any number of photons between $0$ and $n$.

The number of masks reachable from a monomial $\bold{a}$ essentially depends on $d = |\bold{a}| = \sum_{i=1}^m a_i$ and $k$. Any way of distributing between $0$ and $n-d$ photons over the $k$ modes will lead to a different mask, and there are $\binom{n-d+k}{n-d}$ of them. The only exception is the case $k=m$ where only masks with $n$ photons are accepted as there is no mode left. In this case we need to distribute exactly $n-d$ photons over the $k$ modes and there are $\binom{n-d+k-1}{n-d}$ ways of doing it. 

Overall, the time complexity is 

\[ \sum_{\text{monomial $\bold{a}$}} \binom{n-|t|+k-\alpha}{n-|t|} \times 2s_{\bold{a}} \]

with $\alpha=1$ if $k=m$ and $0$ otherwise.

Finally, we iterate over the set of monomials by iterating over the size of the support and the number of photons in the state. For a given support size $s$ and a given number of photons $d$, there are necessarily one photon in each mode of the support (otherwise the support would not be of size $s$) and there are $\binom{d-1}{d-s}$ ways to distribute the remaining $d-s$ photons into the $s$ modes. There are $\binom{m}{s}$ different support of size $s$ possible, such that the final time complexity of the approach is 

\[ \sum_{s=1}^n \sum_{d=s}^{n} 2s \times \binom{n-d+k-(k=m)}{n-d} \times \binom{m}{s} \times \binom{d-1}{d-s}. \]

\subsection{Memory complexity}

As we iterate over the measurement outcomes, and each computation is independent from the others, the memory complexity is given by the maximum memory needed for one call to SLOS with mask. In other words, the memory complexity is 
\[ \max_{\substack{\text{mask $\ket{\mathfrak{m}}$}\\ \text{on $k$ modes}}} \big| \{ \text{state $\ket{t}$ is a parent of any state with mask $\ket{\mathfrak{m}}$} \} \big|. \]

For a given mask, the cardinality is given by
\[ \prod_{i=1}^k (\mathfrak{m}_i+1) \times \binom{m-k+n-p}{n-p} \]

where $p = \sum_{i=1}^k \mathfrak{m}_i$. In other words, given a mask $\ket{\mathfrak{m}}$, the number of parent states is given by the number of submasks $\prod_{i=1}^k (\mathfrak{m}_i+1)$ multiplied by the number of ways we can add at most the remaining $n-p$ photons in the $m-k$ other modes. 

To optimize over the mask $\ket{\mathfrak{m}}$, we first rewrite the memory complexity as 
\[ \max_{p=1\hdots n} \max_{\substack{\text{mask $\ket{\mathfrak{m}}$}\\ \text{on $k$ modes} \\ \text{with $p$ photons}}} \prod_{i=1}^k (\mathfrak{m}_i+1) \times \binom{m-k+n-p}{n-p}. \]

It is known that  

\[ \max_{\substack{\text{mask $\ket{\mathfrak{m}}$}\\ \text{on $k$ modes} \\ \text{with $p$ photons}}} \prod_{i=1}^k (\mathfrak{m}_i+1) \leq \left(1 + \frac{p}{k} \right)^k \]

and for simplicity we assumed that we reach this upper bound. In practice, we saw no difference from actually computing the exact maximum. Therefore, in our setting, the memory complexity of SLOS with masks on $k$ modes is 
\[ \max_{p=1\hdots n} \left(1 + \frac{p}{k} \right)^k \times \binom{m-k+n-p}{n-p}. \]

\end{document}